\documentclass[epj]{webofc}
\usepackage[utf8]{inputenc}
\usepackage[varg]{txfonts}   
\usepackage{booktabs}
\usepackage{xcolor}
\definecolor{darkred}{rgb}{0.4,0.0,0.0}
\definecolor{darkgreen}{rgb}{0.0,0.4,0.0}
\definecolor{darkblue}{rgb}{0.0,0.0,0.4}
\usepackage[bookmarks,linktocpage,colorlinks,
    linkcolor = darkred,
    urlcolor  = darkblue,
    citecolor = darkgreen]{hyperref}

\usepackage{amsmath}
\usepackage{amssymb}
\usepackage{graphicx}
\usepackage{multirow}
\usepackage{braket}
\usepackage{tikz}
\usepackage{tabularx}
\usepackage{diagbox}
\usepackage{booktabs}
\usepackage{subfigure}
\wocname{EPJ Web of Conferences}
\woctitle{Lattice2017}
\begin{document}
%
\selectlanguage{english}
\title{$b\bar b u\bar d$ four-quark systems in the Born-Oppenheimer\\ approximation: prospects and challenges}
\author{%
\firstname{Antje} \lastname{Peters}\inst{1}\fnsep\thanks{Speaker, \email{peters@th.physik.uni-frankfurt.de}}\and
\firstname{Pedro} \lastname{Bicudo}\inst{2} \and
\firstname{Marc}  \lastname{Wagner}\inst{1}
}
\institute{%
Goethe-Universit\"at Frankfurt am Main, Institut f\"ur Theoretische Physik,\\
Max-von-Laue-Stra{\ss}e 1, D-60438 Frankfurt am Main, Germany
\and
CFTP, Departamento de Física, Instituto Superior Técnico, Universidade de Lisboa,\\
Avenida Rovisco Pais, 1049-001 Lisboa, Portugal
}
\abstract{%
We summarize previous work on $\bar b \bar bud$ four-quark systems in the Born-Oppenheimer approximation and discuss first steps towards an extension to the theoretically more challenging $b\bar b u\bar d$ system. Strategies to identify a possibly existing $b\bar b u\bar d$ bound state are discussed and first numerical results are presented.
}
\maketitle

\section{Motivation}
A number of mesons observed in experiment, e.g.\ at the LHCb or at Belle, are tetraquark candidates. Examples are the charged states $Z_b(10610)$ and $Z_b(10650)$, cf.\ e.g.\ \cite{Belle:2011aa}. They are bottomonium-like, which can be concluded from their masses and decay products. However, they also carry electric charge, which means they must include additional quarks. It is most likely that these additional quarks are a light quark-antiquark pair. So the quark content of the $Z_b$ states is assumed to be $b\bar b u\bar d/b\bar bd \bar u$.

\section{Investigating heavy-heavy-light-light four-quark systems with Lattice QCD and within the Born-Oppenheimer approximation}
Four-quark states $b \bar b q\bar q$ are object of experimental and theoretical research. The theoretical investigation of such systems is, for reasons which will be described below, very challenging. Four-quark states $\bar b \bar b qq$ with $q\in\{u,d,s,c,\}$ are theoretically more straightforward to investigate. These states have not yet been measured experimentally. However, it is interesting to study them e.g.\ to be able to make predictions for experiment and to gain conceptual insight of four-quark states. Previous studies of $\bar b\bar b qq$ and $b\bar bq\bar q $ systems can be found e.g.\ in \cite{Stewart:1998hk,Michael:1999nq,Cook:2002am,Doi:2006kx,Detmold:2007wk,
Bali:2010xa,Brown:2012tm, Francis:2016hui,Alberti:2016dru}. 

\begin{figure}[htbp]
\begin{center}
\includegraphics[width=6cm]{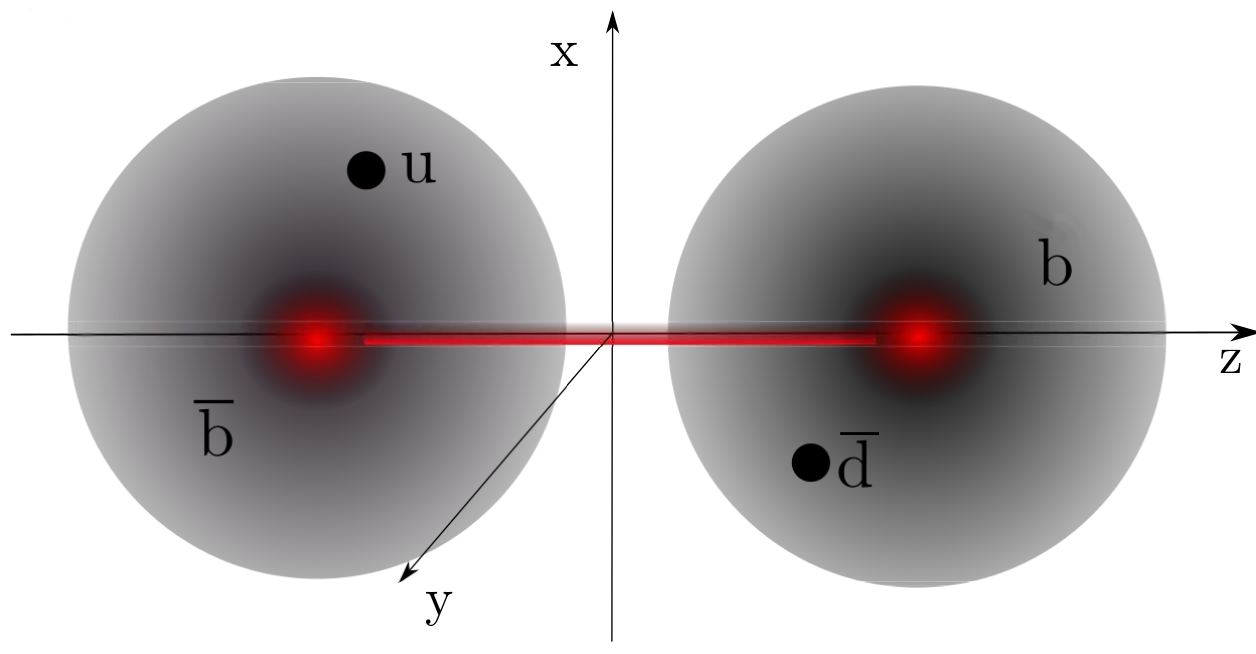}
\caption{The $\bar b \bar b ud$ system for short and long $\bar b \bar b$ separations. (a) At short separations the heavy quarks interact via one-gluon exchange. (b) At large separations, the interaction of the heavy quarks is screened by the lighter quarks and the four quarks form two weakly interacting $B$ mesons.}
\label{BBsketch}
\end{center}
\end{figure}

To investigate binding of four-quarks $\bar b \bar b ud$ and $b\bar b u \bar d$, respectively, we work in the so-called Born-Oppenheimer approximation \cite{BO}. We first consider the $b$ quarks to be static. Consequently, the heavy quark spin decouples from the system. We consider $u/d$ quarks with a finite mass. The central element of the investigation is the potential of the two static quarks in the presence of the lighter quarks which we compute using Lattice QCD. We extract potentials $V(r)$ from correlation functions
\begin{equation}
\label{corr} C(t,r) \ \ = \ \ \bra{\Omega}  O^\dagger(t)  O(0) \ket{\Omega} \ \underset{t \textrm{ large}}{\propto} \ \exp(-V(r) t) .
\end{equation}
$O$ denotes a four-quark creation operator with defined quantum numbers (cf.\ \cite{Wagner:2010ad} for details). It is possible to consider many different flavor/parity/angular momentum channels. For large separations between the static quarks, the potentials correspond to static-light $BB$ potentials and $B\bar B$ potentials, respectively, cf.\ Figure \ref{BBsketch} for the $BB$ case. Once the potentials are computed, we insert them into the Schr\"odinger equation and check for bound states or resonances. 

\section{$\bar b \bar b ud$ systems in the Born-Oppenheimer approximation: recent results}
\begin{figure}[htbp]
\begin{center}
\includegraphics[width=8cm]{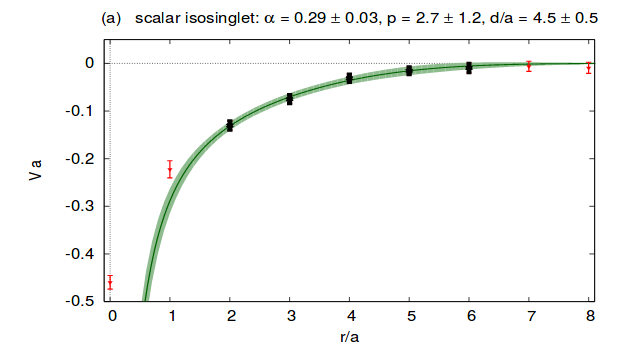}
\caption{The static-light $\bar b \bar b ud$ potential as a function of the $\bar b \bar b$ separation in units of the lattice spacing $a=~0.079$fm.}
\label{BBpotemtial}
\end{center}
\end{figure}

We first consider potentials of two static antiquarks $\bar b \bar b$ in the presence of lighter quarks $ud$. An example plot is shown in Figure \ref{BBpotemtial}. The green line corresponds to a three parameter fit with respect to $\alpha, d$ and $p$ of the function $V(r)=-\frac \alpha r \exp\left(-\left(r/d\right)^p \right)$ to the lattice data, cf.\ \cite{Bicudo:2012qt}. The investigation predicts two tetraquark states that have not yet been measured experimentally: A $\bar b \bar b ud$ bound state in the $I(J^P)=0(1^+)$ channel can be identified. The binding energy with respect to the $m_B+m_{B^*}$ threshold is $E_B=-90^{+43}_{-36}$ MeV \cite{Bicudo:2015kna} not considering effects from the heavy quark spin and $E_B=-59^{+30}_{-38}$ MeV \cite{Bicudo:2016ooe} taking into account heavy quark spin effects. Furthermore, a resonance in the $I(J^P)=0(1^-)$ channel is found. The resonance energy with respect to the $2m_B$ threshold is $E_R=17^{+4}_{-4}$ MeV and the resonance width is $\Gamma=112^{+90}_{-103}$ MeV \cite{Bicudo:2017szl}.

\section{$b\bar b u\bar d$ systems in the Born-Oppenheimer approximation}
The recently measured states $Z_b(10610)$ and $Z_b(10650)$ are experimentally interesting examples for a $b\bar b q \bar q$ four-quark candidate. They have the quantum numbers $I(J^{ P})=1(1^+)$. In the following we consider the positively charged state $Z_b^+$ (cf.\ e.g.\ \cite{Belle:2011aa, Collaboration:2011gja,Ohkoda:2012rj}). However, all results also hold for $Z_b^-$. Isospin $I=1$ corresponding to the positive electric charge is realized by light quark flavours $u\bar d$. Parity $P=+$ is consistent with a possible loosely bound $B^{(*)} \bar B^*$ structure, since both $B^{(*)}$ and $\bar B^*$ have $P = -$ and hence in combination result in $P = +$. Note, however, that in the static approximation $B$ and $B^*$ mesons are degenerate in mass. Therefore we will denote both, $B$ and $B^*$ as $B$ in the following. Numerically we find a rather deep and wide potential for light total angular momentum $j = 0$. In the static approximation the different spin alignments of the static quarks are degenerate, i.e.\ we cannot distinguish $j_b=0$ or $j_b=1$. This means, the total angular momentum can either be $J = 0$ or $J = 1$, i.e.\ all our statements apply to $b\bar b u\bar d$ four-quark system not only with $I(J^{P})=1(1^+)$, but also with $I(J^{P})=1(0^+)$. Up to now, only the $I(J^{ P})=1(1^+)$ channel has been measured experimentally. To measure the $I(J^{P})=1(0^+)$ channel a different experimental setup than realized in current experiments would be necessary.

\subsection{$b\bar b u\bar d$ systems: possible structures}
\label{sec:posssce}
$b\bar b u\bar d$ states in the $I(J^{ P})=1(1^+)$ channel may have different structures. We distinguish between four-quark structures such as the mesonic molecule $B\bar B$ and two-particle states such as a bottomonium state and a pion, $Q\bar Q+\pi$.  Examples of the different and frequently discussed structures and their descriptions can be found in Table \ref{tab:strukturenmitbildern}. 
\begin{table}[hb]
\begin{center}
\begin{tabular}{p{2.5cm} p{8cm}p{3cm}}
label &description &  sketch \\ 
\toprule $B \bar B$\newline{\small tetraquark or\newline two-particle state}& A bound four-quark state made of a loosely bound $B \bar B$ meson pair (a so-called mesonic molecule) or two far separated and essentially non-interacting $B$ mesons&
\begin{tikzpicture}[baseline={([yshift=-10pt]current bounding box.north)}]
    \draw [dashed] (0.7cm,0) ellipse (1.4cm and 0.7cm);
    \draw[thick,red!50!white] (0,0) circle (0.5cm) node {$B$};
    \draw[thick,blue!50!white] (0:1.4cm) circle (0.5cm) node {$\bar B$};
\end{tikzpicture}
 \\ 
\midrule 
   $(Q \bar Q)^ *\pi$ \newline{\small tetraquark or\newline two-particle state}&A bound four-quark state made of an excited bottomonium state and a loosely bound pion $\pi^+$ with zero momentum or two far separated and essentially non-interacting mesons. In the static approximation, a bottomonium state is realized by the static quark $Q$ and the static antiquark $\bar{Q}$ connected by a gluonic string.&  \begin{tikzpicture}[baseline={([yshift=-10pt]current bounding box.north)}]
    \draw [dashed] (0.7cm,0) ellipse (1.4cm and 0.7cm);
    \draw[thick,gray] (0,0) circle (0.5cm) node {$Q\bar Q$};
      \draw[thick,green, fill=white] (0:1.3cm) circle (0.5cm) node {$\pi$};
\end{tikzpicture} \\ 
\midrule  
 $Q \bar Q\pi_{\mathbf p}$ \newline{\small tetraquark or\newline two-particle state}&A bound four-quark state made of the bottomonium ground state $\Upsilon(1S)/\eta_b(1S)$ and a loosely bound pion $\pi^+$ with momentum $\mathbf p$ (including $\mathbf p=0$) or two far separated and essentially non-interacting mesons.&\begin{tikzpicture}[baseline={([yshift=-10pt]current bounding box.north)}]
    \draw [dashed] (0.7cm,0) ellipse (1.4cm and 0.7cm);
    \draw[thick,gray] (0,0) circle (0.5cm) node {$Q\bar Q$};
    \draw[thick,green!80!white, opacity=0.3] (0.2cm:1.1cm) circle (0.5cm) ;
     \draw[thick,green!80!white, opacity=0.6] (0.1cm:1.2cm) circle (0.5cm) ;
      \draw[thick,green, fill=white] (0:1.3cm) circle (0.5cm) node {$\pi_{\mathbf{p}}$};
\end{tikzpicture}\\
\midrule 
{\small tetraquark state}&A bound four-quark state made of a diquark (color antitriplett) and an anti-diquark (color triplett). & \begin{tikzpicture}[baseline={([yshift=-10pt]current bounding box.north)}]
    \draw (0.7cm,0) ellipse (1.4cm and 0.7cm);
    \draw[thick, orange] (0,0) circle (0.5cm) node {$\bar d\bar Q$};
      \draw[thick, yellow, fill=white] (0:1.3cm) circle (0.5cm) node {$u Q$};
\end{tikzpicture}\\
\bottomrule
\end{tabular} 
\end{center}
\caption{Examples for frequently discussed structures of a $b\bar bu\bar d$ state in the $I(J^P)=1(1^+)$ channel. In the text, the structures are referred to by their labels.}
\label{tab:strukturenmitbildern}
\end{table}

\begin{figure}[htbp]
\begin{center}
\includegraphics[width=7cm]{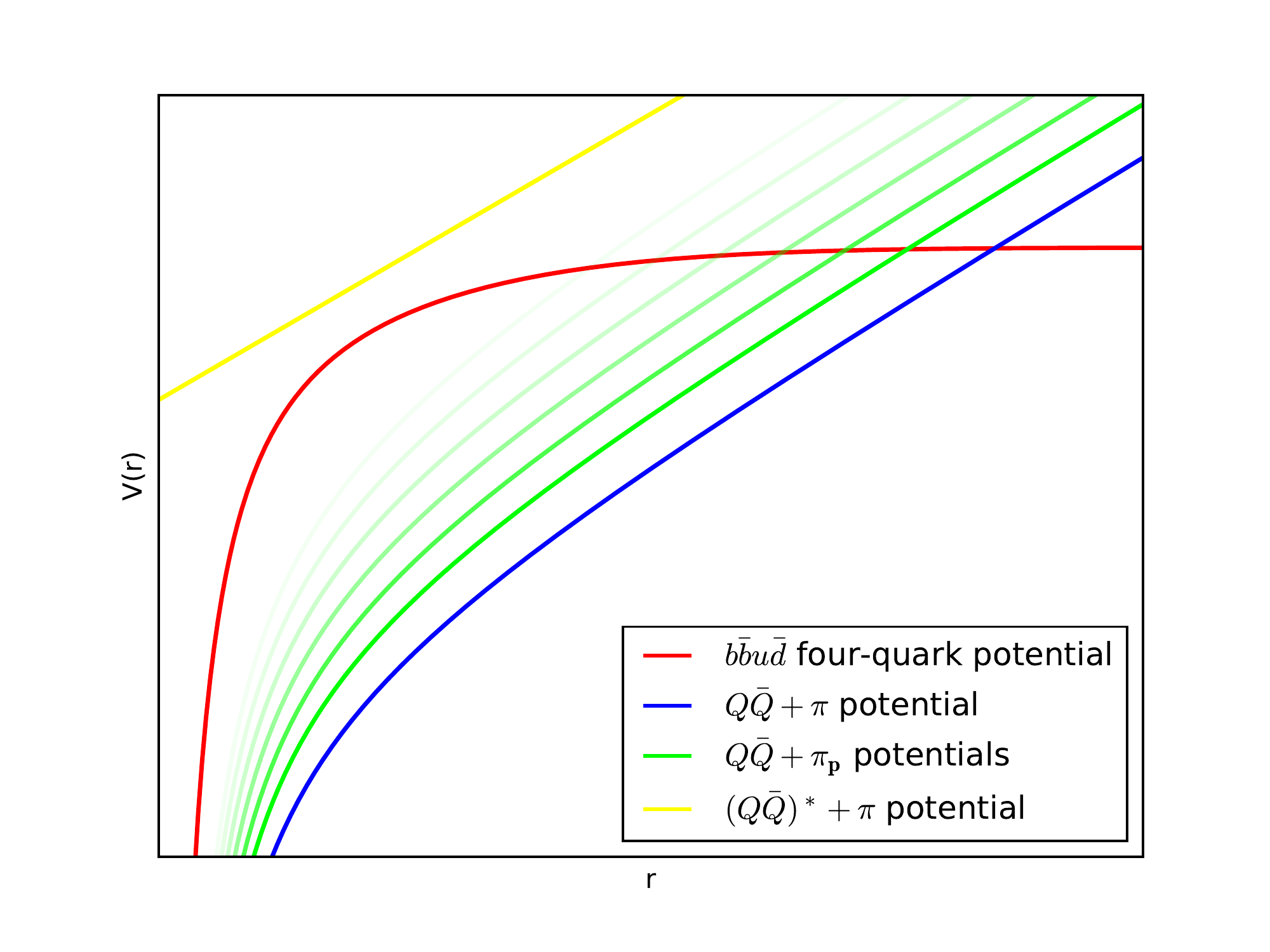}
\caption{Cartoon-like illustration of one possible scenario for the $b\bar b u\bar d$ spectrum. {\color{blue}blue}: Potential of the bottomonium ground state and a pion at rest, $Q \bar Q+\pi$. {\color{green}green}: Green curves indicate $Q \bar Q+\pi_{\mathbf p}$ potentials for different momenta $\mathbf p$. {\color{red}red}: $b\bar b$ potential in the presence of lighter quarks $u\bar d$. {\color{yellow}yellow}: Potential of the excited bottomonium state and a pion at rest, $(Q \bar Q)^*+\pi$.}
  \label{pic:scenario}
  \end{center}
\end{figure}

In the following, we list the potentials that belong to the $b\bar b u \bar d$ ground state and higher excited states in the $I(J^{P})=1(1^+)$ channel.
\begin{itemize}
\item Ground state (denoted as $V_0(r)$):
\begin{itemize}
\item As numerical results indicate (cf.\ Section \ref{groundAndFirst}), one can identify $V_0(r)$ with the bottomonium ground state $\Upsilon(1S)/\eta_b(1S)$ and a pion at zero momentum, $Q \bar Q+\pi$. The bottomonium state is represented by two static quarks connected by a gluonic string. 
\end{itemize} 
\item First excited state (denoted as $V_1(r)$):
\begin{itemize}
\item For small separations $r$ of $b$ and $\bar b$ one can distinguish different cases:
\begin{itemize}
\item A two-particle state $B+ \bar B$ where two mesons are far separated and basically not interacting or a four-quark state $B \bar B$ where four quarks form a hadron,
\item a diquark-antidiquark state,
\item an excited bottomonium state and a pion, realized as a two-particle state $(Q \bar Q)^*+\pi$ where the pion is essentially equally distributed over space or a four-quark state $(Q \bar Q)^*\pi$ where the pion is close to the $Q\bar Q$ pair or
\item a two-particle state corresponding to a bottomonium state and a pion with nonzero momentum $Q \bar Q+\pi_{\mathbf p}$ or a four-quark state $Q \bar Q \pi_{\mathbf p}$.
\end{itemize}
Of course in QCD the state can also correspond to a mixture of the above mentioned structures or a combination of four quarks without any manifest structure. 
\item For large separations $r$ of $b$ and $\bar b$ the first excited state and the ground state swap places (cf.\ Figure \ref{pic:scenario}). The first excited state corresponds to a bottomonium state and a pion at zero momentum. The ground state corresponds to a two-particle state of a $B$ meson and a $\bar B$ meson, $B+ \bar B$. This can be understood as follows: The gluonic string between the two heavy quarks will not persist for large separations, because its energy increases linearly. Therefore all structures different from $B+ \bar B$ are excited states.
\end{itemize}
\end{itemize}
One possible scenario is sketched in Figure \ref{pic:scenario}. The blue curve is the ground state potential $V_0$ which is expected to have the same shape as the static quark-antiquark potential shifted by the mass of the pion. The ordering of the next excitations is, however, less clear. In particular in case of non-vanishing pion momentum, it depends on the light quark mass as well as on the spatial lattice extent $L$: momentum values are quantized on the lattice, $\mathbf p= \frac{2\pi}{L}\mathbf n$ with $n_i=0,1,...,L/a -1$ (with $a$ the lattice spacing). So for larger lattice extents, the potentials that correspond to non-zero momentum pions will move closer and closer together and more states will be below a potential corresponding to a four quark bound state. This behavior is indicated by the fading green curves in the Figure. The red curve accounts for a possible $b\bar bu\bar d$ four-quark potential which one would like to identify in order to investigate a possibly existing tetraquark state. The yellow curve is the potential of the first excitation of a bottomonium state and a pion at rest. It is an open and interesting question whether it is above the four-quark potential or below.

\subsection{Numerical investigation of $b\bar b u\bar d$ ground state and first excited state in the $I(J^P)=1(1^+)$ channel}
\label{groundAndFirst}
\begin{figure}[htbp]
\begin{center}
\includegraphics[width=8cm]{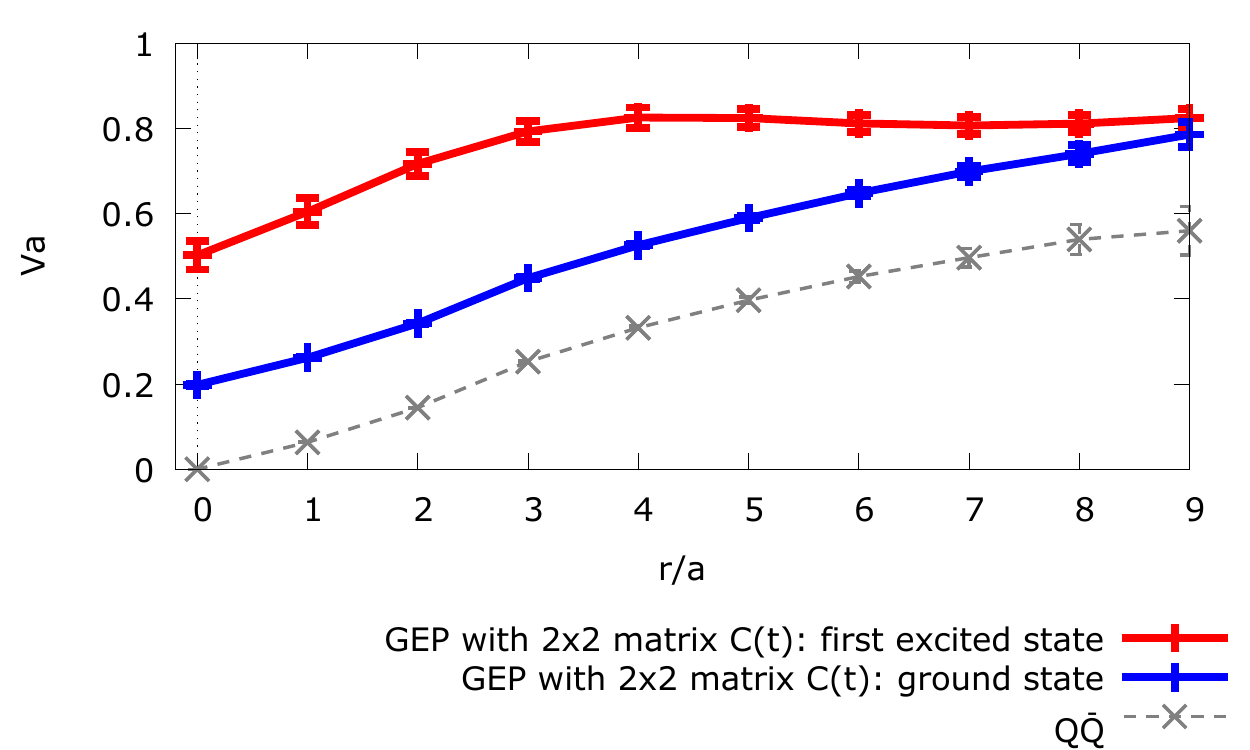}
\caption{The $b\bar b u \bar d$ potentials extracted with the GEP.}
\label{pic:2x2}
\end{center}
\end{figure}

We investigate the ground state and first excited $b\bar b u \bar d$ state by considering creation operators of $Q \bar Q+\pi$ (respectively $Q \bar Q\pi$) and $B\bar B$ (respectively $B+\bar B$) structures. The aim is to check whether the first excited $b\bar b$ potential in the presence of lighter quarks $u\bar d$ is still attractive enough to host a bound state if contributions from the $Q \bar Q+\pi$ state have been removed. 
We compute the correlation matrix $C_{jk}(t,r)$ which can be expressed in terms of contributions of potentials and overlaps $A^n_{jk}$:
\begin{equation}
C_{jk}(t,r)=\bra \Omega O_j^\dagger(t)  O_k(0) \ket \Omega \underset{t\to \infty}{=}A^0_{jk}\exp{(-V_0(r)t)} + A^1_{jk}\exp{(-V_1(r)t)}+...
\label{overlaps}
\end{equation}
with $|\mathbf x - \mathbf y|=r$. The operators read:
\begin{align}
 &O_1= O_{B\bar B}=\Gamma_{AB} \tilde \Gamma_{CD}\bar Q^a_C(\mathbf x) q^{a}_A(\mathbf x) \bar q^{b}_B(\mathbf y) Q	^b_D(\mathbf y)\label{EQN001},\\
 &O_2= O_{Q\bar Q+\pi}= \bar Q^a_A(\mathbf x) U^{ab}(\mathbf x; \mathbf y)\tilde \Gamma_{AB} Q^b_B(\mathbf y)\sum_{\mathbf z} \bar q^{c}_C(\mathbf z)\left(\gamma_5\right)_{CD} q^{c}_D(\mathbf z)\label{QQpi}.
 \end{align}
$\tilde\Gamma$ appearing in both operators is a combination of Dirac matrices that realizes either $j_b=0$ or $j_b=1$. It does not affect the potentials $V_j(r)$ since the spin of the heavy quarks is irrelevant as mentioned above. The matrix $\Gamma$ is a combination of Dirac matrices leading to the same quantum numbers $(j_z , P\circ C , P_x)$ as the $Q \bar Q + \pi$ operator \eqref{QQpi} (for details on ($j_z$, $ P$, $ C$, $P_x$), cf. e.g.\ \cite{Bicudo:2015kna}). Among the possible choices for $\Gamma$, the combination $\Gamma = (1 - \gamma_0) \gamma_5$ yields the strongest $Q \bar Q$ attraction, if one takes into account only the operator $\mathcal O_{B\bar B}$. Therefore, we consider this combination as the most promising to search for a stable $b\bar b u \bar d$ tetraquark. $U^{ab}(\mathbf x; \mathbf y)$ denotes a product of gauge links connecting the two static quarks.
We determine the first excited state potential by solving the Generalized Eigenvalue Problem \cite{Blossier:2009kd}. This potential as well as the ground state potential and the static-light quark-antiquark potential for comparison can be found in Figure \ref{pic:2x2}. Solving the Schr\"odinger equation yields a binding energy of
\begin{equation}
E_B=(-58 \pm 71) \textrm{ MeV}
\end{equation}
with respect to the $m_B+m_{B^*}$ threshold. This value might be a vague indication for a tetraquark state.

\subsection{Future plans to investigate the first excited $b\bar b u\bar d $ state}
To investigate the structure of the first excited state $\ket 1$ we plan to use the overlap of tetraquark and two-particle trial states, respectively, and $\ket 1$. The overlaps can be computed using Equation \eqref{overlaps} and e.g.\ the operators given in Equations \eqref{EQN001} and \eqref{QQpi}. The volume dependence of the overlap can be estimated by a simple quantum mechanical calculation: The static quarks are located at fixed positions $\mathbf x_0$ and $\mathbf y_0$, respectively. The lighter quarks have no fixed location. Their positions are referred to as $\mathbf u$ and $\mathbf v$. The quantum mechanical wavefunctions of the tetraquark and the two-particle trial state each are composed of the wavefunctions of the heavy quarks which are Dirac $\delta$ functions as well as of the wavefunction of the light quarks $\psi_{4q/2p}(\mathbf x)$. The wavefunctions of the heavy quarks decouple from the system because the heavy quark positions are fixed. In the following, we only consider the wavefunctions of the light quarks $\psi_{4q}$ and $\psi_{2p}$. The tetraquark trial state can be modeled as:
 \begin{equation}
\psi_{4q}(\mathbf u, \mathbf v)= f(\mathbf x_0-\mathbf v)f(\mathbf y_0-\mathbf u)
\end{equation}
and the two-particle trial state can be modeled as:
\begin{equation}
\psi_{2p}(\mathbf R, \mathbf r)=\frac 1 {\sqrt V_s} e^{i\mathbf p \cdot \mathbf R} g(\mathbf r)
\end{equation}
where we use center-of-mass coordinates $\mathbf r =\mathbf u - \mathbf v$ and $\mathbf R=\frac{\mathbf u + \mathbf v}{2}$. $\mathbf p$ is the pion momentum. The normalization $\frac 1 {\sqrt V_s}$ makes the state independent of the spatial volume $V_s$. We introduce the functions $f(\mathbf r)$ and $g(\mathbf r)$ that are 0, if $|\mathbf r| >d_{\mathrm{hadron}}$, else nonzero functions (where the details are irrelevant). $d_{\mathrm{hadron}}$ is the typical extent of a hadron, i.e.\ $d_{\mathrm{hadron}}\simeq 1$fm.
We consider the overlap of the trial states with the first excited $b\bar b u\bar d$ state $\ket 1$:\\

\underline{Case 1: $\ket 1$ is a two-particle state:}\\

\begin{equation}
\psi_1=\frac 1 {\sqrt V_s} e^{i\mathbf p \cdot \mathbf R} g'(\mathbf r)
\end{equation}
with $g'(\mathbf r)=0$ if $|\mathbf r| >d_{\mathrm{hadron}}$, else a nonzero function.
\begin{equation}
\braket{ 1| {\psi_{2p}}}= \frac 1 V_s \int d^3 R  \int d^3 re^{-i\mathbf p \cdot \mathbf R} g^{*\prime}(\mathbf r)e^{i\mathbf p \cdot \mathbf R} g(\mathbf r)= const.
\end{equation}
\begin{equation}
\begin{aligned}
\braket{ 1 | {\psi_{4q}}}=&\frac 1{\sqrt V_s} \int d^3 u \int d^3 v e^{-i\mathbf p \cdot \mathbf R} g^{*\prime}(\mathbf r)f(\mathbf x_0-\mathbf v)f(\mathbf y_0-\mathbf u)\\
& \sim \begin{cases}
\frac{1}{\sqrt V_s} & \text{ if } |\mathbf x_0 - \mathbf y_0|<d_{\mathbf{hadron}}\\
0& \text{ otherwise }
\end{cases}.
\end{aligned}
\end{equation}

\underline{Case 2: $\ket 1$ is a tetraquark state:}\\

\begin{equation}
\psi_1=f'(\mathbf x_0-\mathbf v)f'(\mathbf y_0-\mathbf u)
\end{equation}
with $f'(\mathbf r)=0$, if $|\mathbf r| >d_{\mathrm{hadron}}$, else a nonzero function.
\begin{equation}
\begin{aligned}
\braket{ 1| {\psi_{2p}}}=& \frac 1 {\sqrt V}  \int d^3 u \int d^3 v e^{-i\mathbf p \cdot \mathbf R} g(\mathbf r)f^{*\prime}(\mathbf x_0-\mathbf v)f^{*\prime}(\mathbf y_0-\mathbf u)\\
&\sim\begin{cases}
\frac{1}{\sqrt V} & \text{ if } |\mathbf x_0 - \mathbf y_0|<d_{\mathbf{hadron}}\\
0& \text{ otherwise }
\end{cases}.
\end{aligned}
\end{equation}
\begin{equation}
\braket{ 1 | {\psi_{4q}}}= \int d^3 u \int d^3 v f(\mathbf x_0-\mathbf v)f(\mathbf y_0-\mathbf u)f^{*\prime}(\mathbf x_0-\mathbf v)f^{*\prime}(\mathbf y_0-\mathbf u) =const.
\end{equation}
The conclusion is that the volume dependence of the overlap can provide information whether the first excited $b\bar b u\bar d $ state is a two-particle state or a four-quark state (tetraquark). Table \ref{voldeptable} summarizes the results obtained above. More detailed information about the internal structure of the first excited state can be obtained by implementing further operators. 
\begin{table}[hb]
\begin{center}
\begin{tabular}{|c|c|c|}
\hline 
\diagbox{$\ket 1$ is a...}{overlap with...}&  two-particle trial state & tetraquark trial state \\ 
\hline 
tetraquark state & $\sim \frac{1}{\sqrt V_s}$ & no volume dependence \\ 
\hline
two-particle state & no volume dependence & $\sim\frac{1}{\sqrt V_s}$ \\ 
\hline 
\end{tabular} 
\end{center}
\caption{Summary of the volume dependence of overlaps of the first excited four-quark state $\ket 1$ and different trial states.}
\label{voldeptable}
\end{table}

\section{Summary and outlook}
We investigate the static-light $b\bar b u \bar d $ four-quark state in the $I(J^{P})=1(1^+)$ channel. A $b\bar b u \bar d $ bound state must have two properties: The light quarks must be close to the heavy quarks and the corresponding potential must be sufficiently attractive to host a bound state. We take into account different possible structures of the $b\bar b u \bar d $ state and identify a candidate for an attractive $b\bar b u \bar d$ potential. By calculating the corresponding binding energy we find signatures consistent with a $b\bar b u \bar d $ tetraquark. The same methods applied in case of the $Z_b$ states described here could be applied to less well understood states, e.g.\ the X(3872) \cite{Abe:2003hq}, or used to predict new states. However, it is questionable whether it is sensible to treat the $c$ quark within the Born-Oppenheimer approximation. In the future, the first excited $b\bar b u \bar d $ state should be studied in more detail, e.g.\ by investigating the volume dependence of the state with sufficiently large statistics. We present a possible strategy to investigate this volume dependence by means of the overlap of the first excited state with different trial states. This way one can find out whether the first excited state is a tetraquark or a two-particle state. This can be an important step for a solid interpretation of the resulting potential and thus for any statement about the $b\bar b u \bar d $ tetraquark.

\section*{Acknowledgments}

P.B.\ thanks IFT for hospitality and CFTP, grant FCT UID/FIS/00777/2013, for support. M.W.\ and A.P.\ acknowledge support by the Emmy Noether Programme of the DFG (German Research Foundation), grant WA 3000/1-1. 

This work was supported in part by the Helmholtz International Center for FAIR within the framework of the LOEWE program launched by the State of Hesse.

Calculations on the LOEWE-CSC high-performance computer of Goethe-University Frankfurt am Main were conducted for this research. We would like to thank HPC-Hessen, funded by the State Ministry of Higher Education, Research and the Arts, for programming advice.

\bibliography{lattice2017}

\end{document}